\documentclass[preprint,,showpacs,showkeys,preprintnumbers,amsmath,amssymb,MnSymbol]{revtex4}
\usepackage{graphicx} 
\usepackage{bm}
\usepackage[colorlinks=true,urlcolor=red,linkcolor=blue,citecolor=blue]{hyperref}

\newcommand{\ia }{\'{\i}}   
\newcommand{\tetra}{T$_{\rm h}$} 

\newcommand{\octatrun}{t-O$_{\rm h}$}
\newcommand{\cubo}{co-O$_{\rm h}$}
\newcommand{\ico}{m-I$_{\rm h}$} 
\newcommand{\deca}{s-D$_{\rm h}$}
\newcommand{\mar}{m-D$_{\rm h}$}
\newcommand{\ino}{i-D$_{\rm h}$}
\newcommand{\decmon}{dm-D$_{\rm h}$}
\newcommand{\chiu}{c-I$_{\rm h}$}
\newcommand{\icots}{st-I$_{\rm h}$}
\newcommand{\icotd}{dt-I$_{\rm h}$}

\newcommand{\cien}{$\langle 100 \rangle$}
\newcommand{\once}{$\langle 111 \rangle$}
\begin{document}

\title{The Decmon-type Decahedral Motif In Metallic Nanoparticles} 

\author{J.P.\ Palomares-B\'aez$^{1}$, J.L.\ Rodr\ia guez-L\'opez,$^{1,\dagger}$
J.M.\ Montejano-Carrizales,$^{2}$ and  M.\ Jos\'e-Yacam\'an$^{3}$} 
\affiliation{$^1$Advanced Materials Department, IPICYT,
Camino Presa San Jos\'e 2055 78216 San Luis Potos\ia , SLP, Mexico }
\affiliation{$^2$Instituto de F\ia sica,  
Universidad Aut\'onoma de San Luis Potos\ia \-
Alvaro Obreg\'on 64, 78000 San Luis Potos\ia ,  SLP, Mexico }
\affiliation{$^3$Department of Physics and Astronomy
and  International Center for Nanotechnology and  Advanced Materials (ICNAM)\\ 
 University of  Texas At San Antonio 78249-1644,~San Antonio Texas, USA }

\begin{abstract}
Structural and energy stability results for a new class of 
decahedral structural motif  termed {\it decmon} (Montejano's decahedra) are presented.\  
After making proper truncations  to the regular icosahedron,
 this structural motif presents exposed   \cien \- and \once \- facets,   
with an energy competition that makes the structures very stable.\ 
A structural transition as a function of the cluster size is also identified.  
An outline for a path transformation from the Mackay icosahedra (\ico) to the regular decahedra 
(\deca) symmetry structures, as well as
experimental evidence of the decmon decahedral motif in metallic nanoparticles are presented.\ 
\end{abstract}



\maketitle

In the 1960's,  Shozo Ino was the first to report a  
very complete study on the structural phases
for supported  multi-twinned gold nanoparticles (NPs)--tetrahedra (\tetra), truncated
cuboctahedra (\cubo), Mackay icosahedra (\ico), and regular decahedra
(\deca)~\cite{Ino69}, by developing a 
theory that accounted for the specific surface,  twin
boundaries, elastic strain, and the adhesive energies to the substrate.
This theory came after extensive studies of
epitaxial growth of fcc metals on rocksalt faces by   
  Ino~\cite{Ino66}
and Ino and Ogawa~\cite{Ino67}, where they proposed 
the {\em multi-twinned particle model with a nucleus of (001) orientation},
now known as the Ino decahedral (\ino) family.\ 
The Ino decahedron  is a truncated decahedron with lower 
total surface-to-volume ratio  that exposes higher
energy \cien \- facets parallel to the five-fold axis, and
 is energetically more stable than the simple (bi-pyramid) decahedra (\deca)
but not with respect to other multi-twinned nanoparticles (MTNPs) like the 
\ico \- in the small and intermediate size range ($\approx$ 10 nm)~\cite{Ino69}.\ 
 
This description was prevalent for around a  decade or so, and although
 the   thermodynamic processes of shape and morphology formation 
for fcc particles  was well
understood by then using of the 
Wulff construction~\cite{Wulff01} the more general problem
of twinned nanoparticles was not.\  With the help  
of a modified Wulff  construction for multi-twinned particles, 
  Marks~\cite{Marks83}     proposed  a 
modification to the Ino decahedra  which allowed for
nonconvex re-entrant facets at the twin boundaries of the
decahedron.\ This structure is now known  as the Marks Decahedron 
(\mar) and is energetically competitive with the \ico \- in the small-size
 range, and even more stable than other multi-twinned 
particles such as the \deca \-, the  \ino \- and 
the \ico \- structures in the medium and large size range.\ %

Recently,    Barnard and
coworkers~\cite{Barnard08} have applied  nonconvex re-entrant features
to the regular Mackay-Icosahedron (\ico) to obtain the so-called 
Chui-Icosahedron (\chiu), a modified icosahedron that, in the smallest 
re-entrant reconstruction, each particle  contains 12 atoms fewer than the regular 
\ico \- family members.\ This type of reconstruction had been observed 
in experiments with decahedral nanoparticles by 
Rodr\ia guez-L\'opez {\em et al.}~\cite{JLRDZ04}, and in many 
reports by molecular dynamics (MD) simulations~\cite{Chiu06} (see also
inside Ref.~\onlinecite{Barnard08}).\   
 
Another related structure has been observed by Ascencio {\it et al.}~\cite{Ascencio00} 
by means of high resolution transmission electron microscopy
(HRTEM) characterization.\ They observed
images with a contrast similar to the icosahedral or the truncated
decahedra in gold nanoparticle samples.\ However, they also observed
pseudo-square faces of type \cien \-  together with triangular faces \once \-;\ 
therefore they proposed a new structure termed the {\it truncated icosahedron} (t-I$_{\rm h}$).\ 
 
Thus, non-crystallographic atomic arrangements, such as icosahedral (\ico) and decahedral (\deca) 
symmetries and some variations of them have been widely established, both from 
atomistic simulations~\cite{Cleveland97} and first-principle calculations~\cite{Balleto02}
 in the small-size range  (1-2 nm)   
  or  experimentally~\cite{Marks94,Martin96,Koga04}, and even other MTNPs like
the bi-\ico~\cite{Marks83,Koga06} have been observed; 
these non-crystallographic symmetry structures are lower in energy  than the fcc pristine
structures, such as the truncated octahedra (\octatrun) and the cubo-octahedra (\cubo).\  
A plausible explanation for this fact is
that the observed MTNPs  are in metastable states but with a lower
free energy barrier from the liquid to the \ico \- phase compared to the
barrier from the liquid to the fcc crystalline phase~\cite{Nam05}.\  
 
In this letter we present two new sub-families   
 that result from 
particular truncations  made to the regular icosahedron.\ 
This type of truncation exposes  facets \cien \- and  
\once-- in addition to the external \once \- facets in the 
regular icosahedron--that  improve the 
energy stability of these new kind of decahedral nanoparticles compared with the 
icosahedron near that size.\

FIGURE 1 HERE

\begin{figure}
\begin{center}
\includegraphics[width=\linewidth]{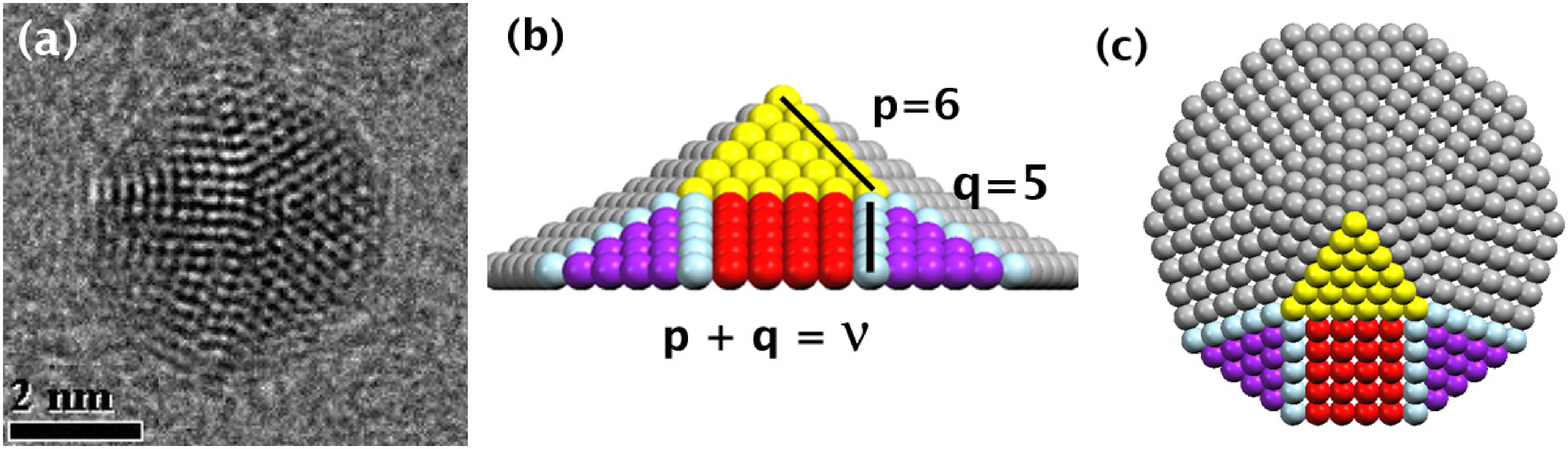}
\caption{(a) Experimental image that shows a contrast following 
  the pyramidal decmon-type decahedral  growth pattern.\ 
The Montejano's Decahedral ({\em decmon})  motif shown in side and top views, 
a decahedral atop motif that is shared by the single  (\icots) and double  (\icotd) truncated icosahedra, 
as well   as  the decmon subfamily  (\decmon).\ Indexes ($p,q$) are shown for this model.} 
\label{decepx}
\end{center}
\end{figure}

In Fig.~\ref{decepx}(a), we present a sample of a HRTEM image of
a gold nanoparticle, where  a central pentagonal pyramid arrangement can be observed, 
as well as some \once \- and \cien \- arrays.\ After these intriguing
experimental observations, 
we introduce a model to reproduce the images observed.\ The
model shown in Figs.~\ref{decepx}(b) and \ref{decepx}(c), is formed by a central pentagonal 
pyramid of triangular faces, where each one of these faces is connected to \cien-type array and
each vertex to \once-type array and these two arrays are join
together into couples.\ 
This model is  a pyramid termed the {\it decmon} (Montejano's decahedron) motif, 
and it presents a decagonal base
along with  a pentagonal pyramid at the top.\ 
 
\begin{figure}
\begin{center} 
\includegraphics[width=0.7\linewidth]{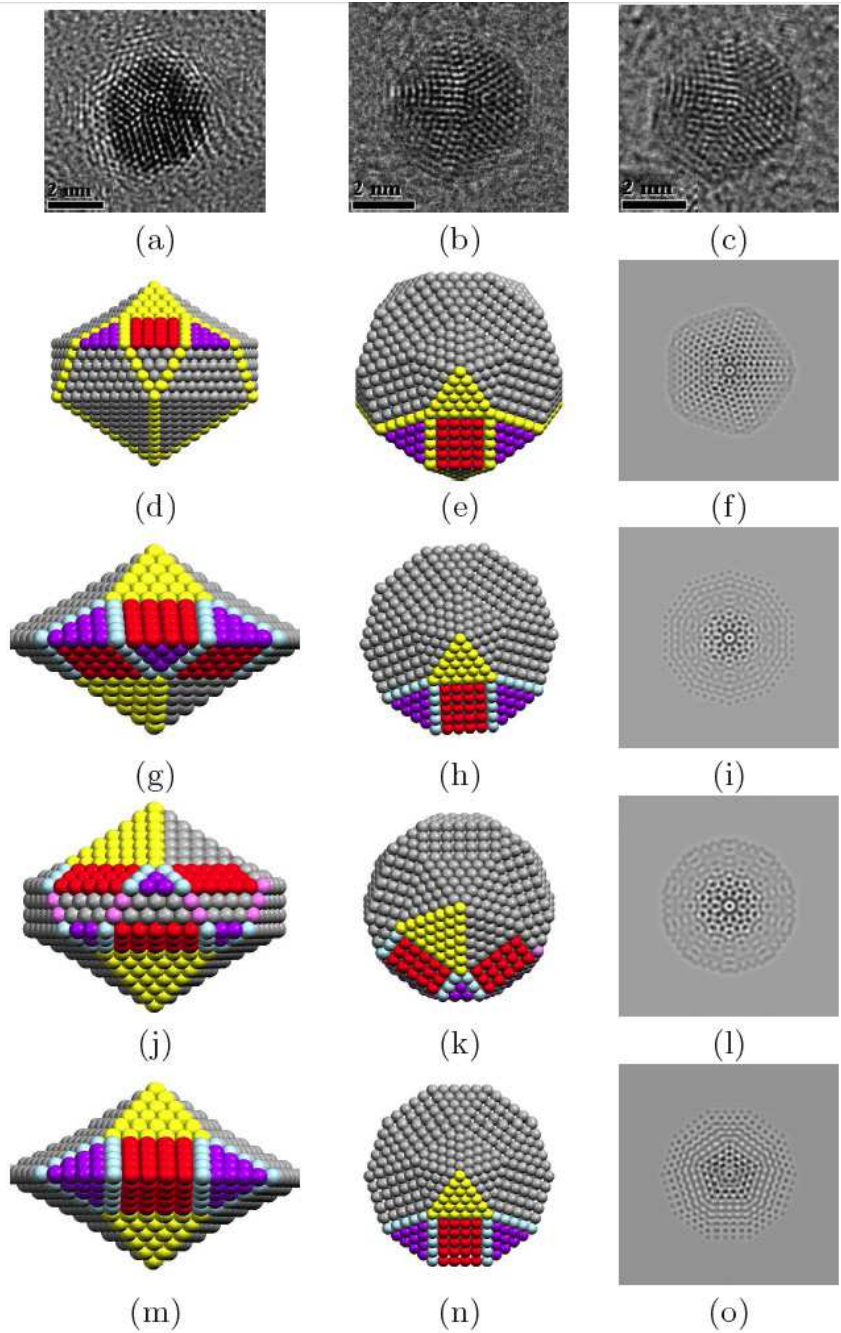}
\caption{From (a)-(c), the HRTEM experimental images show a contrast following the
pyramidal decmon-type decahedral growth pattern, 
contrasting them with simulations of HRTEM  (E = 200 kV, Cs = 0.5 mm 
at optimal defocus Sherzer)~\cite{simulatem};
the best fitting can be seen in (f).\    
A comparison of representative members of the Montejano's 
decahedra family is shown for  $^5$\icots$^{11}$ \- (d-e), 
the  $^5$\icotd$^{11}$ \- (g-h) and   $^3$\icotd$^{10}$ \- (j-k), 
and the regular \decmon (m-n).\ 
Models are colored to show structural differences among them.}
\label{todos}
\end{center}
\end{figure}

The model proposed in Fig.~\ref{decepx} is only the
top view of the particle and full particles could adopt one of the
four different shapes seen in Fig.~\ref{todos}.\ 
These truncated icosahedra come from particular truncations made to the regular icosahedron, 
and the truncations are made  drawing out (first) one pentagonal 
cap to a given $\nu$ order icosahedron 
(the order $\nu$ is defined as the number of atoms in one edge, including both vertices), 
resulting  a single truncated icosahedron 
($^{q=1}$\icots$^\nu$,  where $q$ is termed the order of the truncation 
made to the \ico, and the following relation $\nu  = p + q$ holds).\  
If a second truncation is made to a single truncated icosahedron, the top cap to be
eliminated is different from the one discussed above;\ now the top cap
of the $^q$\icots$^\nu$ \- is formed by all the atoms of the decmon motif, 
{\em but the number of atoms in both top caps is the same}, whether the 
cap has been truncated or not.\
With this fact in mind, successive truncations can be made either in the
 $^q$\icots$^\nu$ \- or in the  $^q$\icotd$^\nu$.
 There are single (Figs.~\ref{todos}(d)--\ref{todos}(e)) and  double (Figs.~\ref{todos}(g)--\ref{todos}(k)) 
truncations  made to the $\nu=11$ and 10 order icosahedra, respectively.\  
All these structures form a new set of nanoparticle sub-families members 
with decahedral symmetry, the so-called {\it decmon} family.\

FIGURE 2 HERE

 Experimental results on gold NPs are shown in 
Figs.~\ref{todos}(a)--\ref{todos}(c), where   
 a central five-fold  pyramid with  some 
\cien \- and  \once \- atomic arrays is observed.\   
These intriguing HRTEM image contrasts 
are compared with simulations using representative 
members of the decmon family (see the third column in  Fig.~\ref{todos}).\ 
From these results, the model that best resembles the experimental 
contrast is in Fig.~\ref{todos}(f), which comes from  the single truncated icosahedron proposed by 
Ascencio {\it et al.}~\cite{Ascencio00}.\ 
Models are shown for the  $^5$\icots$^{11}$  \-  
in Figs.~\ref{todos}(d)--\ref{todos}(f); 
the   double truncated icosahedra family  is shown in  
  two types  ( $^5$\icotd$^{11}$ \-  in Figs.~\ref{todos}(g)--\ref{todos}(h)   
and  $^3$\icotd$^{10}$ 
 in Figs.~\ref{todos}(j)--\ref{todos}(k), 
where $\nu$ is the order of the icosahedron which they came from).\ 
Finally,  in  Figs.~\ref{todos}(j)--\ref{todos}(l),   
the regular decmon structure (\decmon) is also shown.
  
A regular icosahedron of order $\nu$ is a cluster with a central site and $\nu-1$
concentric icosahedra shells; observed perpendicularly to the five-fold symmetry axis, 
it has pentagonal pyramids  at the top and the bottom. 
This picture contrasts with  Fig.~\ref{todos}(d) where the \icots \-
has a decmon pyramid (similar to Figs.~\ref{decepx}(b)), at the top and a pentagonal
one at the bottom.\ Furthermore, when the  top caps of two opposite vertices are
eliminated from an icosahedron, the double truncated icosahedron is
obtained (\icotd), shown in  Figs.~\ref{todos}(g) and 
\ref{todos}(j), with  a decmon pyramid at each side.\

For the single truncation case in a $\nu$-order icosahedron,  
there are $\nu-1$ single
possible truncations; the shape  for any $^q$\icots$^\nu$ \- is
a decmon pyramid on one side and a pentagonal one in the opposite vertex.\  
Once all the possible truncations have been made, 
a fully single truncated icosahedron 
is obtained with an  structure which is an irregular decahedron.\

Regarding the double truncations permissible in a
$\nu$-order icosahedron, there are $[\nu-1]/2$ $([\nu-1]/2)$ for $\nu$ (even
or odd).\ The shape for any \icotd \- is formed by two
opposite decmon pyramids joined by 10 trapezoidal
lateral faces.\ However, the final shape for the fully $^q$\icotd$^\nu$ \-
 does depend on whether  the order ($\nu$) of the icosahedron is odd 
 (Fig.~\ref{todos}(g))  or even (Fig.~\ref{todos}(j)).\ 

The {\it decmon} type polyhedron, Figs.~\ref{todos}(m)--\ref{todos}(n), 
is a structure that results from reflecting the decmon pyramid 
with respect to the base, being thus a symmetric polyhedron 
with respect to the equator, contrasting with the $^5$\icotd$^{11}$ \-
(Figs.~\ref{todos}(g)--\ref{todos}(h)), which,  despite looking similar, 
represent  very different structural models.\

\begin{figure}
\begin{center} 
\includegraphics[width=\linewidth]{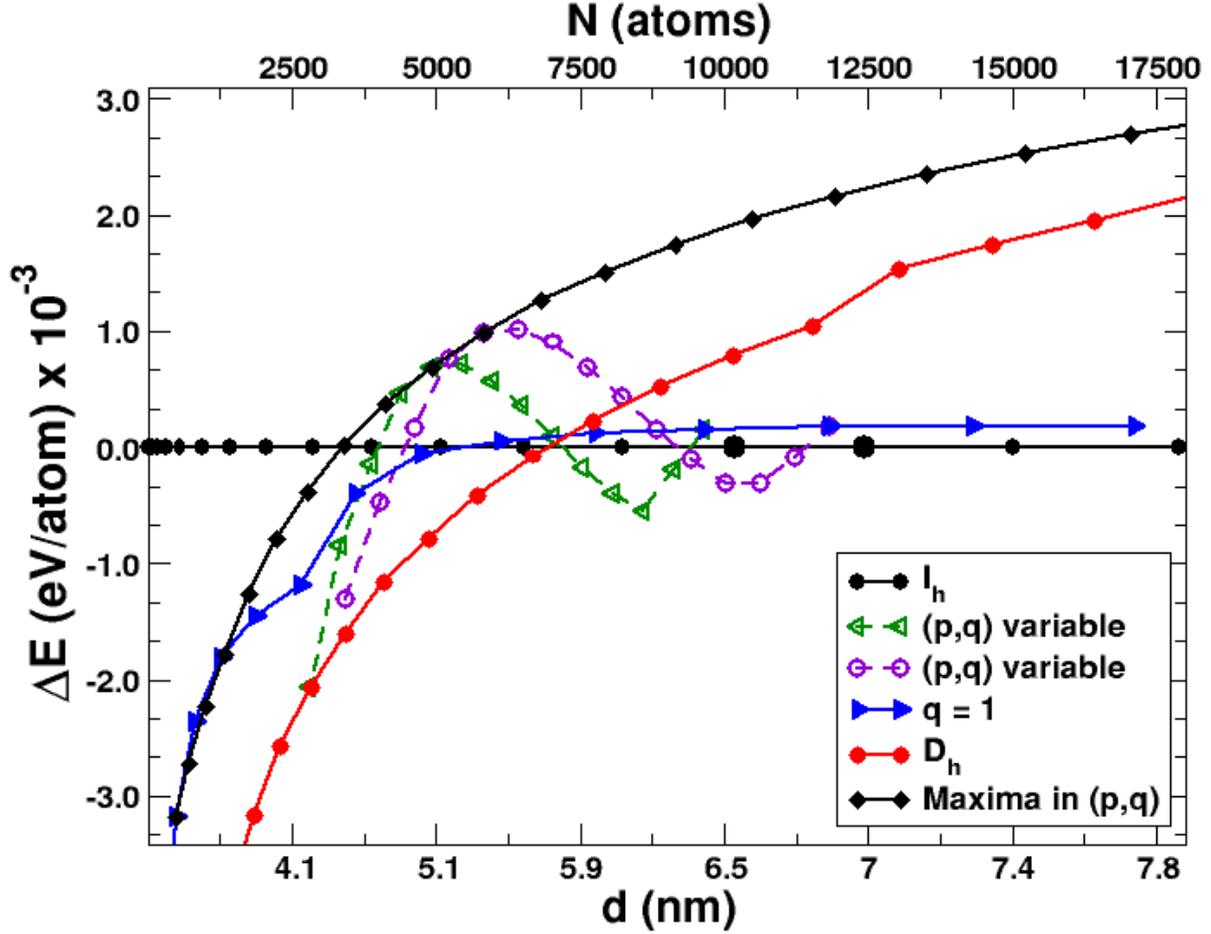}
\caption{Color on-line.\ Different truncations made to the regular icosahedra are shown, 
plotting  energy difference {\em vs.} mean diameter of the particle ($d\propto N^{1/3}$), 
where icosahedra energy is used as reference~\cite{Note1}.
Blue line reflects how the first truncation ($q=1$, blue solid triangles) 
affects the energy of the particles, stabilizing it after a given cluster size.
Big circles shown in the reference line ($\Delta$\;E\,=\,0) 
correspond to Mackay I$_{\rm h}$ of order $\nu\,=\,15$ and 16.\ 
From these clusters, successive single truncations 
(green triangle  and purple circle) make the structures metastable
until  a maximum stability (black diamonds) is obtained.\ 
From these maximum points, there is
a steady decay towards the regular \deca \- (red circles) and other 
closely related decahedral structures; {\em i.e.,} a surface reconstructed decahedron ($\circ$).\    
These truncation paths offer evidence of a structural transformation from the \ico \- to the \deca \- symmetry 
structures.\ }
\label{trunc}
\end{center}
\end{figure}

As previously discussed, the decmon (\decmon) structural motif  
results from a particular 
truncation made to the regular icosahedron, being very 
 different from the one known as the Chui truncation~\cite{Barnard08}.\ 
This is a very efficient way to optimize the energy stability of metallic NPs, 
as can be observed in Fig.~\ref{trunc}, where we use as energy reference
the cohesive energy per atom for the icosahedral family~\cite{Calc}.\ 

FIGURE 3 HERE

First, as seen in Fig. \ref{decepx}(b) the ($p,q$) indexes  define a given \icots, 
we can vary $p$ as a function of  
the number of atoms (N) and $q$ can be kept constant, or viceversa.\ But we can also vary ($p,q$) 
as a function of the number of atoms (N).\     
Keeping $q$ constant in all the size range, 
it should be observed that the first truncation ($q=1$, blue triangles) 
made to all the cluster sizes, 
begins with a positive high slope in the small-size region. 
After a given cluster size (around $d=4.1$ nm), this truncation  improves  
their stability significantly making it even more stable than the icosahedra near $d=5.3$ nm 
($N\approx 5000$ atoms).\ This change in the curvature is size dependent and 
is only  observed  up to the truncation $q=3$ (not shown).\  After that, successive  truncations 
only increase  monotonically  the energy stability of the $^q$\icots.\ 
This abrupt change in the curvature 
of the stability has been identified as a surface reconstruction in the \icots~\cite{Palomares09}.\

Now, if successive truncations are made to a  given icosahedron; 
{\em e.g.} $\nu=15$, with  $ N=10\,179$ atoms, 
and a particle diameter of  $d\approx 9.3$ 
the indexes  ($p,q$) would be changing (keeping the relation $p+q =\nu$) and what is  
obtained is a  transformation path from the \ico \- to the 
\deca \- symmetry structures (green $\lhd$ in Fig. \ref{trunc}, 
where each point represents cohesive energy per atom for  
a different  $^q$\icots \- structure, but all of them  come from 
the same \ico$^{\nu=15}$).\ 
The first truncation slightly improves its energy,  the next four truncations take the particle 
below the icosahedral energy reference, and then  the next five truncations or structures are more
stable than those  icosahedra   around the respective \icots \- size cluster, reaching a maximum 
in  $\nu-q =5=p$.\ If  truncations are made until they are exhausted, 
an irregular decahedron is obtained, which 
after relaxation turns into either a {\em structure with the same energy for a perfect decahedra} 
(here shown the complete \deca \- family with red circles), 
or a {\em surface reconstructed} decahedra, {\em e.g.}, last truncation 
for \icots$^{\nu=15}$ ($\circ$ in purple).\  

In the size region shown in Fig~\ref{trunc}, $p\,=\,5$ 
for the two path truncations made from the \ico \- to the \deca \- structures  
($\lhd$ from $\nu\,=\,15$ and $\circ$ from $\nu\,=\,16$ icosahedrons, respectively), and 
  is  constant over a given size range, {\em i.e.}  
it is dependent on the cluster size.\  
Therefore, and very interestingly,  there are (size dependent) constants $p_1\,<\, p_2\,<\,\ldots$, 
a fact that is related to  a delicate  competition between energy release strain from the appearance of    
\cien \- {\em vs.}  \once \- facets    in these single truncated icosahedra.\  
All these facts are discussed in more detail in a forthcoming paper~\cite{Palomares09}.\ 

\begin{figure}
\begin{center}
\includegraphics[width=\linewidth]{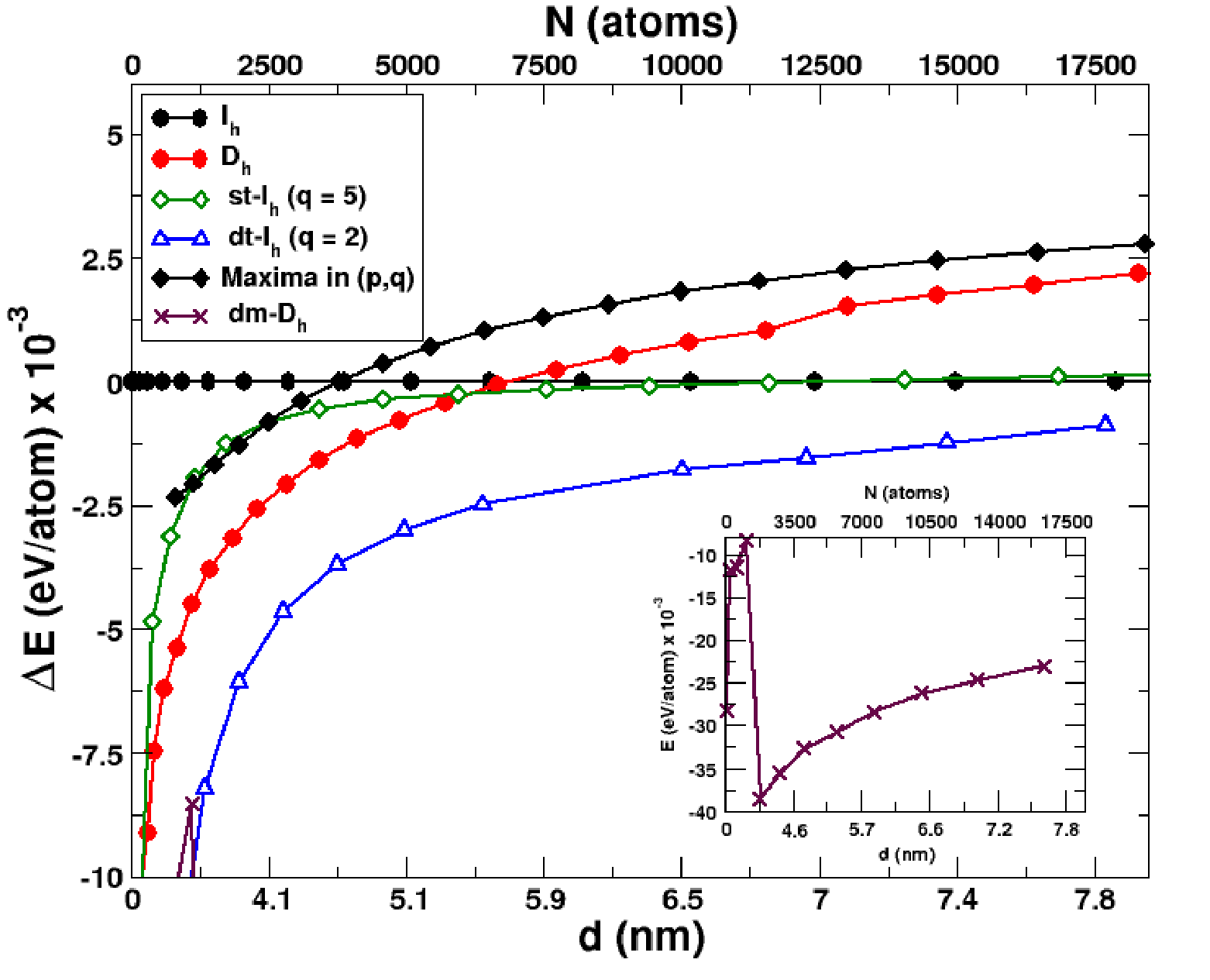}
\caption{Color on-line.\ 
Energy competition between the different sub-families that show the decmon decahedral motif,
plotting  the energy difference {\em vs.} mean diameter of the particle ($d\propto N^{1/3}$), 
where icosahedra energy is used as reference~\cite{Note1}.\ For reference, the curves 
for regular \deca \- (red circles), 
and the maximum obtained with the single truncations (black solid diamonds) are plotted.\ 
Green diamonds  correspond to $^5$\icots \-
and  blue triangles  is for the $^2$\icotd.\ The {\em decmon} (\decmon) structure does not 
compete in energy with these $^q$\icots \- and $^q$\icotd \- structures, 
that is shown in the inset (brown $\times$ symbol).}
\label{eam-fam}
\end{center}
\end{figure}

Figure~\ref{eam-fam} shows  how these decmon subfamilies (\icots, \icotd, and \decmon) compete
among themselves, where the cohesive energy for gold NPs is plotted as a function of the cluster size 
(upper $x$-axis) or as a function of the relative particle diameter (lower $x$-axis).\  
We choose to plot representative truncations for each subfamily, {\em i.e.}, 
the 5$th$ truncation for the \icots \- (green diamonds), 
the 2$nd$ truncation for the \icotd \-(blue triangles), 
the maxima for all the path truncations from the \ico \- to the \deca \- symmetries 
(black solid diamond)   
and in the inset, the energetically non-competitive \decmon \- structure family (brown $\times$).
For  energy reference, the cohesive energies per atom for the icosahedral 
(\ico, black solid circles) as well as  the regular decahedral family 
(\deca, red solid circles) are plotted.\  
  
FIGURE 4 HERE

It can be  concluded from this figure, that up to a given truncation, 
these single truncations made to the regular icosahedron improve 
 its energy stability on the resulting \icots,  but that double truncations do not 
improve the energy of the NPs, as seen for the \icotd \- and  \decmon.\ 
However,  the {\em decmon}  structural motif that  
results from this particular  truncation made to the regular icosahedron  
is a very efficient way to optimize the energy stability of metallic NPs.\

In conclusion, we have introduced the {\em decmon} decahedral motif for metallic NPs, 
which identifies a new family of decahedral structures, 
that after proper truncations made to the icosahedron 
present exposed   \cien    \- and \once \- facets,   
with an energy competition that  makes the structures very 
favorable.\ 
Other outlined aspects worthy of mention, are 
the finding of structural transitions as a function of the cluster size, 
the appearance and competition of surface reconstruction with faceting,   
and the outlined path transformation from the \ico \- to the \deca \- symmetry structures.\ 
Also, we have presented
experimental evidence of the decmon decahedral motif in metallic nanoparticles.\

\section*{Acknowledgments}

The authors acknowledge computing time on the National Supercomputer Center (CNS-IPICYT, Mexico) and the 
 Texas Advanced Computer
Center (TACC)  at UT-Austin, USA.\ Also financial support from
CONACYT grants J42645-F, 50650, 
PIFI P/CA-16 2007-24-21, the Welch Foundation with grant  number  AX-1615 and  the 
National Science Foundation (NSF)  project number DMR-0830074.\ 

$\dagger$~Corresponding author:\ jlrdz@ipicyt.edu.mx

\def\jour#1#2#3#4{{\it#1}\ {\bf#2},\ #3\ (#4)}

\pagebreak
\newpage

\begin{thebibliography}{99}
\bibitem{Calc} All the calculations were performed using the semi-empirical embedded atom method (EAM) as coded 
in the XMD algorithm~\cite{XMD}.\ The clusters were quenched at a rate of $5\times10^{12}$ K/s for 35 ps.\
\end{thebibliography}
\end{document}